\documentclass{emulateapj}

\newcommand{\mincir}{\raise
-2.truept\hbox{\rlap{\hbox{$\sim$}}\raise5.truept\hbox{$<$}\ }}
\newcommand{\magcir}{\raise
-2.truept\hbox{\rlap{\hbox{$\sim$}}\raise5.truept\hbox{$>$}\ }}
\newcommand{\minmag}{\raise
-2.truept\hbox{\rlap{\hbox{$<$}}\raise6.truept\hbox{$<$}\ }}

\shorttitle{AEGIS: the environment of X-ray sources at $z\approx1$}
\shortauthors{Georgakakis et al.}

\begin{document}

\title{All-wavelength Extended Groth strip International Survey: the
  environment of X-ray sources at $z\approx1$}   

\author{
A. Georgakakis\altaffilmark{1,2}, 
K. Nandra\altaffilmark{1}, 
E. S. Laird\altaffilmark{3}, 
M. C. Cooper\altaffilmark{4}, 
B. F. Gerke\altaffilmark{5}, 
J. A. Newman\altaffilmark{6}, 
D. J. Croton\altaffilmark{4}, 
M. Davis\altaffilmark{4,5}, 
S. M. Faber\altaffilmark{7}, 
A. L. Coil\altaffilmark{8}
}
\altaffiltext{1}{Astrophysics Group, Blackett Laboratory, Imperial
  College, Prince Consort Rd , London SW7 2BZ, UK} 

\altaffiltext{2}{Marie-Curie fellow; email: age@imperial.ac.uk}

\altaffiltext{3}{UCO/Lick Observatory, University of California, Santa
  Cruz, CA  95064, USA}

\altaffiltext{4}{Department of Astronomy, University of California at
  Berkeley, Mail Code 3411, Berkeley, CA 94720, USA} 

\altaffiltext{5}{Department of Physics, University of California, 
Berkeley, Mail Code 7300, CA 94720, USA}

\altaffiltext{6}{Hubble fellow; Lawrence Berkeley National Laboratory,
  1 Cyclotron Rd, Mail Stop 50-208, Berkeley, CA 94720, USA} 

\altaffiltext{7}{UCO/Lick Observatory and Department of Astronomy and
  Astrophysics, University of California, Santa Cruz, CA
  95064, USA}

 \altaffiltext{8}{Hubble fellow; Steward Observatory, University of
   Arizona, 933 N. Cherry Ave.,   Tucson, AZ 85721-0065, USA}

\begin{abstract}
We explore the environment of $z \approx 1$ AGN using a sample of 53
spectroscopically identified X-ray sources in the All-wavelength
Extended Groth strip International Survey. We quantify the
local density in the vicinity of an X-ray source by measuring the
projected surface density of spectroscopically identified optical
galaxies within a radius defined by the 3rd nearest neighbour. Our
main result is that X-ray selected AGN at $z \approx 1$ avoid
underdense regions at the 99.89\% confidence level. Moreover,
although we find that the overall population shares the same (rich)
environment with optical galaxies of the similar $U-B$ and $M_B$, there
is also tentative evidence (96\%) that AGN with blue colors ($U-B \la
1$) reside in denser environments compared to optical galaxies. 
We argue that the results above are a  consequence of the whereabouts
of massive galaxies, capable of hosting supermassive black holes at
their centers, with available cold gas reservoirs, the fuel for AGN 
activity. At $z\approx1$ an increasing fraction of such systems are
found in dense regions.      
\end{abstract}


\section{Introduction}\label{sec_intro}
In recent years there has been increasing evidence that the formation
of spheroids and the build-up of supermassive black holes at their
centers are strongly interconnected (Ferrarese \& Merritt 2000;
Gebhardt et al. 2000; Alexander et al. 2005). Moreover, it is now
well established that galaxy properties, such as morphology, color
and star-formation, strongly depend on  environment (e.g. Butcher \&
Oemler 1978; Lewis et al. 2002; Gomez et al. 2003; Hogg et al. 2004),
suggesting a close link between local density and the evolution of
individual systems. Putting the evidence above together, it is natural
to assert that AGN activity, being strongly coupled to galaxy
formation and evolution, should also depend on environment. 

Despite significant observational progress however, the link
between local density and AGN remains controversial. For example, at 
low redshift ($z\approx0.1$) Miller et al. (2003) found no dependence
on environment of the fraction of spectroscopically identified AGN in
the Sloan Digital Sky Survey (SDSS; Schneider et. al. 2005). More
recent studies,  also using SDSS data, suggest that it 
is only when the AGN population is split into subsamples based on optical
classification and/or luminosity, that environmental differences
become apparent.  For example powerful AGN ($L[{\rm O\,III}] > 10^{7}
\, L_{\odot}$) and/or narrow-line Seyferts are found in increasingly 
{\it less} dense regions at $z \approx 0.1$, while less luminous AGN
and/or LINERs show no dependence on local density (e.g. Kauffmann  et
al. 2004; Wake et al. 2005; Constantin \& Vogeley 2006). The above 
low-$z$ results however, appear to be in conflict with observations 
suggesting that at least certain classes of powerful AGN, such as
radio galaxies, reside in relatively rich environments (e.g. Zirbel 
1997). Additionally, the large scale distribution of 
optically and X-ray selected AGN at $z\ga 1$ is consistent with
correlation lengths in the range $r_0 = 5 - 10 \, h^{-1}  \rm \,
Mpc$ (e.g. Croom et al. 2005; Basilakos et al. 2004; Gilli et
al. 2005; Adelberger \& Steidel 2005). This indicates that AGN have
local density distribution similar to early-type systems at $z \approx
1$ ($r_0 \approx \rm 6.6 \,h^{-1}\,Mpc$; Coil et al. 2004b) and that
they avoid poor environments at these redshifts (e.g. emission-line
galaxies, $r_0  \approx \rm 3.2 \,h^{-1}\,Mpc$; Coil et
al. 2004b). Contrary to these results, Coil et al. (2006) show that
the clustering amplitude of broad-line QSOs at $0.7<z<1.4$ matches
that of blue galaxies rather than early-type systems.  This may
indicate differences in the methods used to select AGN in these
studies.

In this paper we directly quantify, for the first time, the environment
of X-ray selected AGN at $z \approx 1$ in an attempt to shed light on
the AGN/density relation at high redshift and how it compares with
low-$z$ results. The testbed for this analysis is the All-wavelength
Extended Groth strip International Survey (AEGIS), a unique
multiwavelength wide-area (0.5\,deg$^2$) sample, which combines
observations from X-ray to radio with moderate resolution optical
spectroscopy to $z \approx 1.4$ (Davis et al. 2006). Throughout the
paper we adopt $H_{0}=\rm 70\,km\,s^{-1}\,Mpc^{-1}$,  $\Omega_{M}=0.3$
and  $\Omega_{\Lambda}=0.7$.

\section{The Data}\label{data}
The main source of redshift information for the AEGIS is the DEEP2, 
a spectroscopic survey which aims to explore the galaxy  
properties and the large-scale structure at $z\approx1$ (Davis et al. 
2003). This on-going project uses the DEIMOS spectrograph on the 
10\,m Keck II telescope to obtain redshifts for galaxies to 
$R_{AB}=24.1$\,mag. The spectra are obtained with a moderately 
high resolution grating ($R \approx 5000$), which provides a 
velocity accuracy of $\approx 30 \rm \, km \, s^{-1}$ and a 
wavelength range  $\rm 6500-9100\,\AA$. This spectral window
allows the identification of the  O\,II emission line to $z  \approx
1.4$. In this study we use a total of 5864 galaxies with secure
redshifts ($>$90\% confidence; quality $Q>2$) in the range
$0.6<z<1.4$. In this interval the AEGIS has high spectroscopic
sampling rate and is less affected by the survey edges.

At X-ray wavelengths the 0.5\,deg$^2$ region of the AEGIS is covered
by 8 ACIS-I {\it Chandra} pointings. In the analysis presented in this
paper we use data from 4 of the 8 {\it Chandra} fields that will
eventually be available for the AEGIS. Details about the X-ray data
used here are listed in Table \ref{tab1}. The remaining observations
are still being analyzed or lie in regions where the DEEP2
spectroscopy is still incomplete and does not allow study of the local 
density using 3-dimensional information. The X-ray data reduction,
source detection and flux estimation are carried out using methods
fully described in Nandra et al. (2005). The detected sources are
optically and spectroscopically identified using the DEEP2 photometric
and redshift catalogues (Coil et al. 2004a; Faber et al. 2006)
following the prescriptions presented by Georgakakis et
al. (2006). The X-ray sample comprises a total of 814 sources of which
58 in the range $0.6<z<1.4$.

\begin{deluxetable}{ccccc}
\tablecolumns{5} 
\tablewidth{0pc} 
\tablecaption{Log of {\it Chandra} observations used in this paper.\label{tab1}}
\tablehead{ 
\colhead{field}    &  \colhead{Observation}   &  \colhead{$\alpha$}  &
\colhead{$\delta$}  & \colhead{exposure} \\
\colhead{name}    &  \colhead{ID}   &  \multicolumn{2}{c}{(J2000)}&
\colhead{time (ks)} \\ 
}\startdata 
 GWS         &  3305, 4357, 4365 & 14:17:43.60 & +52:28:41.20 & 200 \\
 EGS3        &  5845, 5846   & 14:21:32.70 & +53:13:27.70 & 98 \\  	
 EGS4        &  5847, 5848   & 14:19:23.90 & +52:50:32.69 & 89 \\
 EGS7        &  6222, 6223   & 14:15:22.50 & +52:08:26.40 & 84 \\
\enddata 
\tablecomments{Columns are: (1) the name of the {\it Chandra}
  pointing, (2) the sequence number of the observations used at a
  given pointing, (3) nominal right ascension of satellite pointing,
  (4) nominal declination, (5) total exposure time of observations 
   used in this paper.}
\end{deluxetable}

\section{Overdensity estimator}\label{method}
We quantify the environment of galaxies following the prescription
described by Cooper et al. (2005, 2006). For a given source the 
DEEP2 redshift information is used to estimate $D_3$, the projected 
distance on the sky to the 3rd nearest neighbor within a radial velocity 
slice of $\rm \pm1000\,km\,s^{-1}$ from the central source. This  
is then converted to projected density using the relation 
$\Sigma_3= 3 / (\pi \, D_3^2 )$.  We attempt to account for edge
effects by excluding from the analysis sources with projected distance
$<1$\,Mpc from field edges. This reduces the sample size to 4189
optical galaxies and 53 X-ray sources. The $\Sigma_3$ estimates are
then corrected for the DEEP2 variable spectroscopic incompleteness
across the survey region by normalizing with the 2-dimensional
completeness map, which accounts for the redshift identification
success rate at different sky positions (Cooper et
al. 2006). Additionally, DEEP2 is a magnitude limited spectroscopic 
survey and therefore,  $\Sigma_3$ underestimates the true density at 
higher redshifts. We account for this effect by dividing the measured
$\Sigma_3$ at a given $z$ with the median at that redshift (Cooper et
al. 2005, 2006). This normalized  quantity, $\delta_3$, is a measure
of the overdensity in the vicinity of a galaxy relative to the median
field density at that redshift. 

\begin{figure}
\epsscale{0.8} \plotone{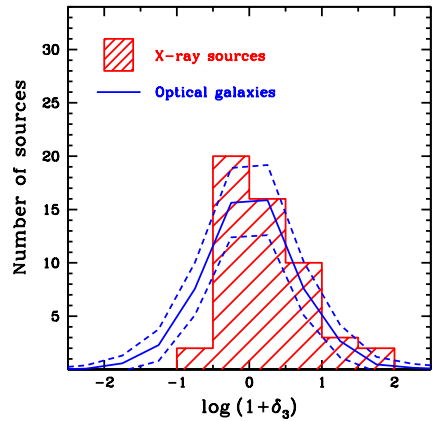}
\caption{
Distribution of $\log (1+\delta_3)$ for the X-ray selected AGN (hatched
histogram). This is compared with the mean expectation for control
samples of optical galaxies (continuous line). The dashed lines
represent the $1\sigma$ rms around the mean.  
 }\label{hist_den.fig}
\end{figure}

\begin{figure*}
\epsscale{0.8}\plottwo{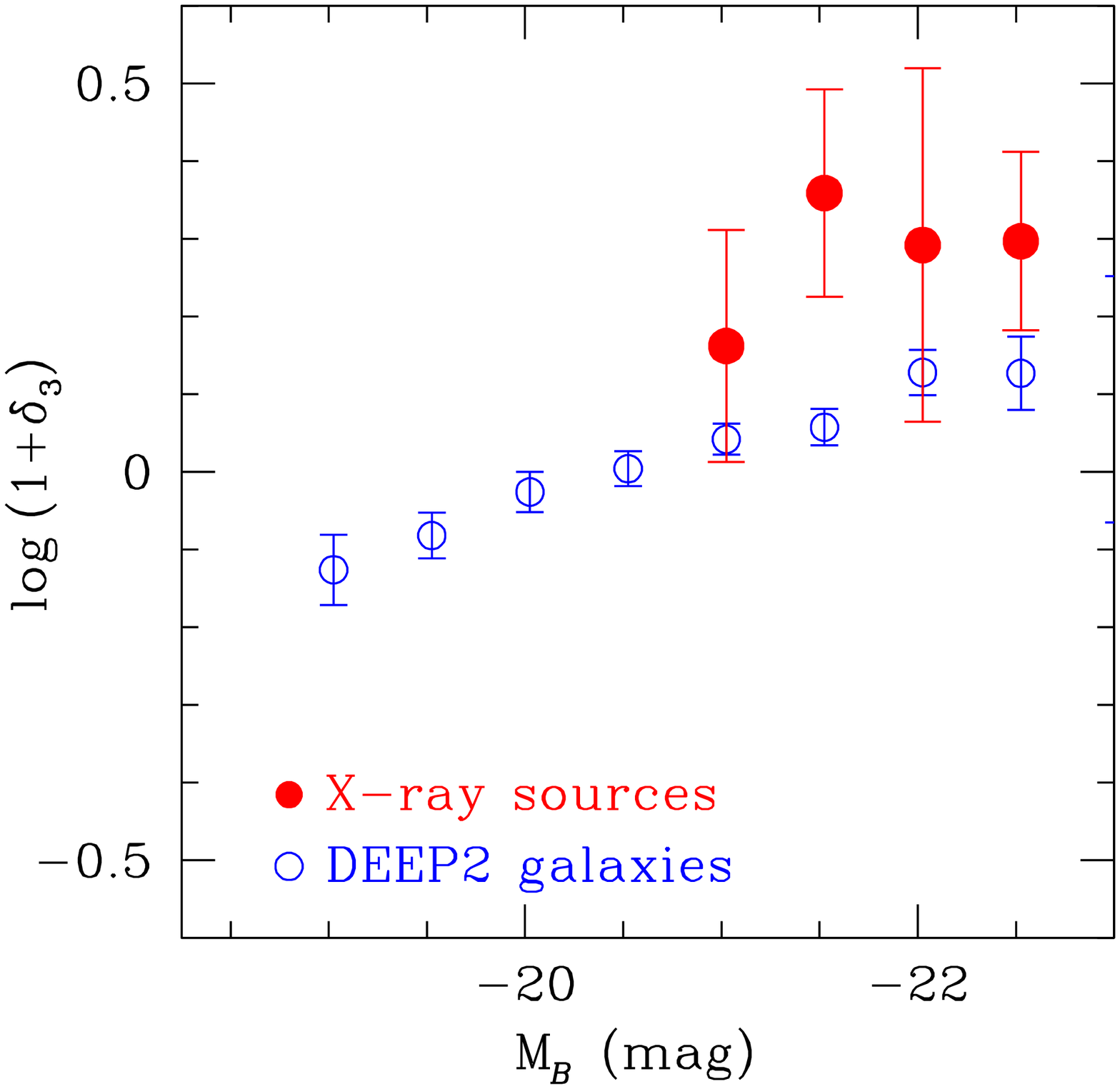}{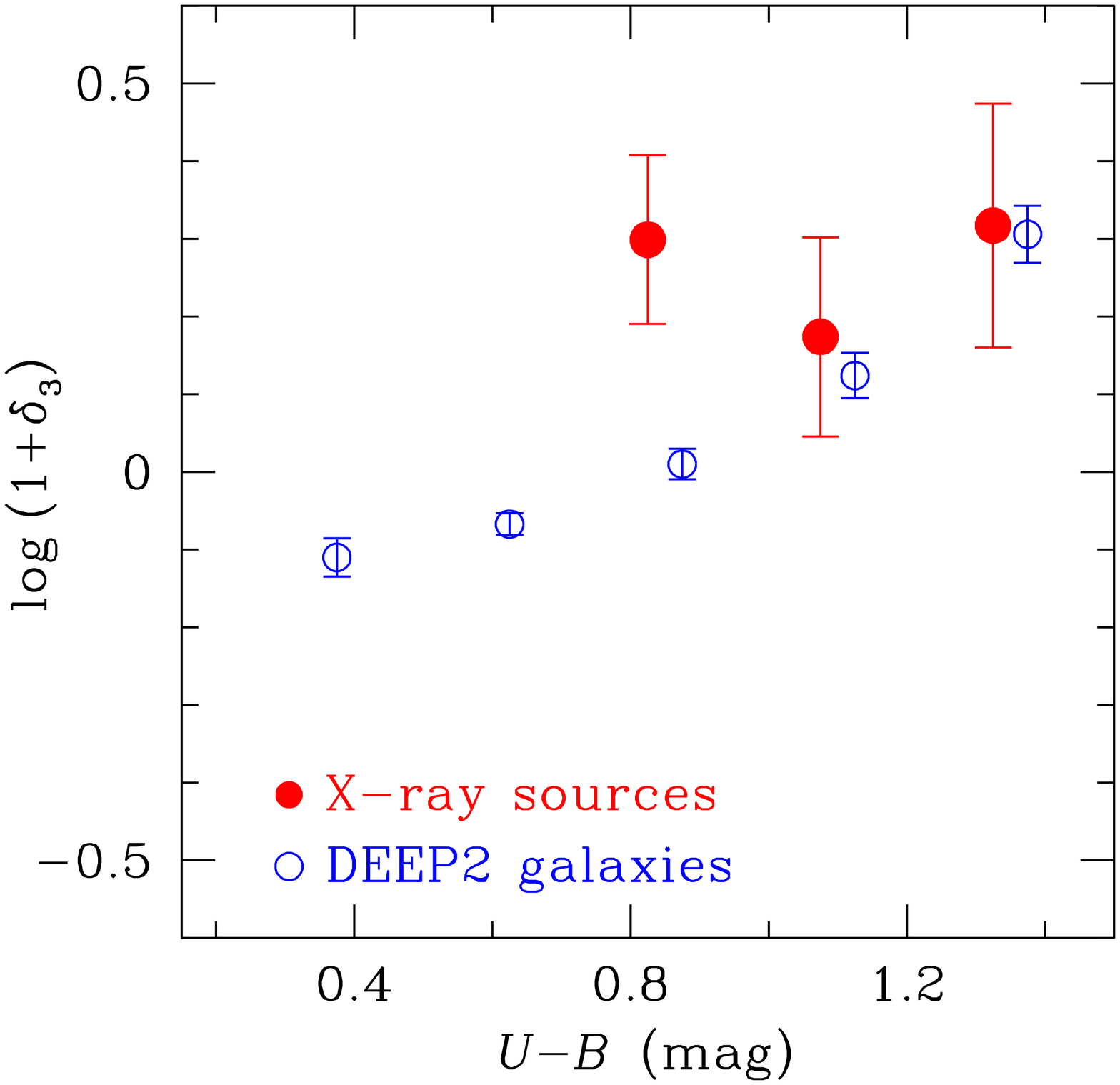}
\caption{
Mean overdensity estimator $\log (1+\delta_3)$ against $M_B$ (left)
and $U-B$ (right). In both plots the mean $\log (1+\delta_3)$ is
estimated in different $M_B$ and  $U-B$ bins. DEEP2 galaxies and X-ray
sources are plotted with open and filled circles respectively. The
errorbars correspond to the standard error in each bin. We only plot
systems brighter than $M_B=-23$\,mag because more luminous galaxies
are affected by small number statistics. For the X-ray sample we only
plot bins with more than 4 sources. The X-ray source points in the 
right panel are offset to the left for clarity. 
 }\label{hist_con_den_color.fig}
\end{figure*}

\section{Results}\label{results}

Fig. \ref{hist_den.fig} presents the distribution of the overdensity
estimator for the X-ray sample in comparison with the mean expectation 
for optically selected galaxies. For the comparison, a total of 200 
control subsamples are constructed by randomly selecting galaxies 
from the DEEP2 spectroscopic survey. The mean $1+\delta_3$ distribution 
for the above 200 subsamples and the $1\sigma$ rms are also shown in 
Fig. \ref{hist_den.fig}. There is evidence in this plot that X-ray 
sources avoid underdense environments, $\log (1+\delta_3) \la -0.5$. 
Indeed, the KS test rejects the null hypothesis probability, $P_{KS}$,
that the two distributions (optical and X-ray) are drawn from the same 
population at the 99.89\% confidence level. Optically luminous systems 
however, have on average different local  environments compared to 
fainter galaxies (e.g. Cooper et al. 2006). Matching the absolute 
magnitude distribution of the random subsamples to that of the X-ray 
sources reduces $P_{KS}$ to 99.4\%. The rest-frame color of galaxies
also strongly depends on environment. We therefore repeat the analysis
by matching  both the $M_B$ and $U-B$ distributions of optical galaxies 
and X-ray sources. This further reduces $P_{KS}$ to 87\%.

The trends above are further explored in
Fig. \ref{hist_con_den_color.fig}, plotting the mean
$\log(1+\delta_3$) within different $M_B$ and $U-B$ bins for both
DEEP2 galaxies and X-ray selected  AGN. When estimating averages we
account for the Malmquist bias by weighting each galaxy by $\rm
1/Vmax$, the inverse of the maximum volume available to a source
within the survey limits (e.g. Cooper et
al. 2006). Fig. \ref{hist_con_den_color.fig} shows that firstly,
X-ray selected AGN are found, on average, in  overdense regions and
secondly, they are associated with optically luminous ($M_B \la
-21$\,mag) galaxies with $U-B \ga 0.7$, in agreement with previous
studies  (Dunlop et al. 2003; Colbert et al. 2005; Grogin et al. 2005;
Nandra et al. 2006). Interestingly though, the two panels of
Fig. \ref{hist_con_den_color.fig} appear to give somewhat different
results on the local density of X-ray sources  relative to optical
galaxies: for a given $M_B$, X-ray sources reside in richer
environments compared to optical galaxies at the 99\% significance
level, while this is not the case when grouped in $U-B$ color, with the
exception of relatively blue AGN ($U-B\approx0.8$). 

\section{Discussion}\label{discussion}

In this paper we explore the environment of X-ray selected AGN at
$z\approx1$ using data from the AEGIS. The main result from our
analysis is that the X-ray population at $z\approx1$ avoids underdense
regions at the 99.89\% significance level. To further understand the
individual trends of environment with color and absolute magnitude we
use the color-magnitude diagram (CMD) of X-ray sources
presented by Nandra  et al. (2006). From this perspective, the
higher mean density of the X-ray sample relative to DEEP2 galaxies at
a given $M_B$ is  because the optical sample includes sources with
very blue colors, known to reside in environments with below average
local density (Fig. \ref{hist_con_den_color.fig}; Cooper et
al. 2006). 

Similarly, a given $U-B$ color bin samples optical galaxies with a
range of absolute magnitudes, with the less luminous ones residing in
underdense environments. The arguments above are consistent with our
finding that the overdensity estimator distribution of X-ray sources
and optical galaxies differ only at the 87\% level when the two
samples are matched in both the $M_B$ and $U-B$. X-ray sources with
relatively blue colors however, appear to be genuinely different from
the DEEP2 population, albeit at the 96\% level. These systems stand
out in Fig. \ref{hist_con_den_color.fig} and this difference persists
at a similar significance, if we restrict the optical sample to
$M_B<-21$\,mag, i.e. to galaxies with similar brightness to X-ray
sources. These AGN occupy the upper bound of the blue cloud in the CMD 
presented by Nandra et al. (2006). Their blue colors may be due to
contamination of the optical continuum by AGN emission. However, out
of the 19 X-ray sources in that color bin only 3 have $L_X > 5 \times
10^{43} \rm \, erg \, s^{-1}$ and therefore may be dominated by AGN
emission at optical wavebands. Alternatively the  above tentative
trend with color  may suggest  that for these systems environment
plays a role in the observed  activity. A larger X-ray sample is
required to further explore this.  

The SDSS spectroscopic survey currently provides the only direct
estimate of the environment of AGN, albeit at much lower redshift,
$z\approx0.1$, compared to this study. Kauffmann et al. (2004) 
using the 1st release of the SDSS found a strong dependence on environment
for powerful  AGN ($L {\rm [OIII]} > 10^7 L_{\odot}$) at $z<0.1$, in
the sense that  the most luminous systems are found in the
field. Lower luminosity AGN in this study show no dependence on local
density. The results above on the environment of low-$z$ AGN are
clearly different from those reported here. Kauffmann et
al. (2003) also explored the 
host galaxy properties of powerful AGN in the SDSS. They find that
they are associated with massive ($ M \ga 10^{10}\, M_{\odot}$) early
type galaxies which are however, distinct from the bulk of the
optically selected early-type population, in that  they show evidence
for on-going or recent star-formation activity. Based on the evidence
above Kauffmann et al. (2003) suggest that there are two essential
ingredients for strong AGN activity: a massive central black hole and
abundant gas supply to fuel it. Only massive early-type galaxies have
large enough bulges to host massive black holes. From these early-type
galaxies those which show evidence for substantial amounts of young
stellar populations have sufficient gas supply to both produce young
stars and to feed the central engine. Such galaxies are relatively
rare in the present day Universe and are preferentially found in
low-density regions. Only in these environments can gas-rich
star-forming galaxies survive today (e.g. Kauffman et al. 2004; Gomez
et al. 2003; Lewis et al. 2002; Poggianti et al. 2006).   

At higher redshifts however, there is an increasing fraction of
massive galaxies with sufficient cold gas reservoirs that can
potentially produce young stars and also fuel luminous AGN. For
example, there is accumulating evidence that the number density of 
luminous blue galaxies, which are rare locally (e.g. Kauffman et
al. 2004), substantially increases to $z\approx1$ (Bell et al. 2004;
Cooper et al. 2006). These systems have blue colors, most likely
because of star-formation, while their luminosities suggest high
stellar masses. These galaxies are clearly prime candidates for
powerful AGN hosts. Interestingly, these luminous blue galaxies,
contrary to their low-$z$ counterparts, are shown to reside in regions
of enhanced density (Cooper et al. 2006). As demonstrated in
Fig. \ref{hist_con_den_color.fig}  our X-ray selected AGN sample is
likely to include such systems. Moreover, the bluer X-ray sources in
the sample appear to reside in higher density regions compared to
optical galaxies on average.

The evidence above suggests that our finding for an association
between X-ray selected AGN and higher density regions at $z \approx 1$
is related to the whereabouts of massive galaxies with available cold
gas reservoirs to sustain accretion of material on the central black
hole. Observations show that such systems are also found in denser
environments at $z\approx1$, contrary to the local Universe. 

A central question in AGN studies is the triggering mechanism of the
observed activity. Recent results suggest that mergers are not the
main process for activating a supermassive black hole (Grogin et
al. 2005; Pierce et al. 2006). Is it possible then, that the
environment plays a role in 
triggering the AGN activity suggesting a causal link between the two?
The enhanced density of the bluer X-ray sources in our sample compared 
to optical galaxies, if confirmed with larger samples, supports such
an association. For example it is plausible that the central black hole
becomes active during the infall of the host galaxy to the overdense
region, as it experiences the gravitational potential of the structure
or frequent interactions with other galaxies in that region. In the
framework of hierarchical models massive galaxies, which can
potentially host luminous AGN, are more often found in denser
regions. At $z \approx 1$, richer environments were substantially more
active compared to the local Universe (Bundy et al. 2006), with the
higher contrast occurring for intermediate-mass  group-like systems
with velocity dispersions $\sigma \approx 500-600 \rm \,  km \,
s^{-1}$ (Poggianti et al. 2006), close to the upper mass limit that is
well sampled by the DEEP2 (Gerke et al. 2005). According to Poggianti
et al., this enhanced activity is associated with infalling systems
many of which are likely to have sufficient mass to harbour a central
black hole and cold-gas reservoirs to sustain star-formation activity
and  possibly also feed  the black hole. As these massive galaxies
further grow (e.g. by  mergers) from high redshift to the 
present day, the supply of cold gas is cut-off (e.g. Croton et
al. 2005; Cooper et al. 2006) leading to quenching of the
star-formation and possibly the AGN activity. These massive  systems 
will therefore, appear as red-and-dead at low-$z$.

A possible link between AGN triggering and environment, if
confirmed, would suggest that the enhanced local density  in the
vicinity of X-ray sources at $z \approx 1$ is a consequence of  the 
hierarchical evolution of structures in the Universe and its impact on
the cold gas reservoirs of individual galaxies. The AEGIS survey
provides a unique dataset that can potentially test this scenario. Open
questions include: what fraction of the  X-ray  population is found in
optically selected groups (e.g. Gerke et al. 2005)? Are there 
morphological and/or color gradients with distance from the centre 
of the group, indicating
enhanced activity for infalling members? Are there differences 
in the color, stellar mass and/or X-ray luminosity between AGN locked
in groups and those that are not? Addressing these points requires a
larger X-ray sample than that presented here. A full discussion of
these issues is therefore  referred to a future publication using the
full AEGIS X-ray sample. 
\\
\\ 
 Financial support has been provided through PPARC and the Marie-Curie
 Fellowship grant MEIF-CT-2005-025108 (AG) the Leverhulme trust
 (KN), the  Hubble Fellowship grants 
 HF-01165.01-A (JAN) and HF-01182.01-A (ALC), the NSF grants AST00-71198 
 and AST0071048.  The W.M. Keck Observatory, a scientific partnership 
 among Caltech, the University of California and NASA. The Observatory
 was made possible by the generous financial support of the W.M. Keck
 Foundation. The authors wish to acknowledge the very significant
 cultural role that the summit of Mauna Kea has within the indigenous
 Hawaiian community; we are fortunate to be able to conduct
 observations from this mountain.

\end{document}